\documentclass[12pt]{iopart}
\usepackage{graphicx}
\def\pmb#1{\setbox0=\hbox{$#1$}%
  \kern-.025em\copy0\kern-\wd0
  \kern.05em\copy0\kern-\wd0
  \kern-.025em\raise.0433em\box0}
\def\parb{\pmb{\partial}}
\def\alt{\mathrel{\hbox{\rlap{\hbox{\lower4pt\hbox{$\sim$}}}\hbox{$<$}}}}
\begin{document}
\title{Numerical simulations of general gravitational singularities}
\author{David Garfinkle}
\address{Dept. of Physics, Oakland University,
Rochester, MI 48309, USA}
\ead{garfinkl@oakland.edu}
\begin{abstract}
This paper covers some of the current techniques and issues involed in
performing numerical simulations of the formation of singularities.
\end{abstract}
\maketitle
\section{Introduction}

A longstanding issue in general relativity has been to find the general
behavior of the singularities that form in gravitational collapse.  
As far back as 1971, a conjecture on this
subject was formulated by Belinskii, Khalatnikov and Lifshitz (BKL).\cite{bkl}
However, BKL provided no means of verifying their conjecture.  To remedy
this situation, Berger and Moncrief and their 
collaborators\cite{bv1,bv2,bd,bjm1,bjm2} in the 1990s 
embarked on a program
of examining the nature of singularities by numerical simulations: choose
initial data whose evolution will become singular and evolve it numerically
to see whether the singularity that forms is of the type conjectured by BKL.

As is often the case in numerical relativity, the Berger-Moncrief plan was
to first treat spacetimes with symmetry and then eventually proceed to the
general case of spacetimes with no symmetry.  Cases with symmetry are often
easier to treat numerically and require a great deal less in the way of
computational resources (or equivalently, with a fixed amount of computational
resources assuming symmetry allows one to obtain much more resolution).  
However, spacetimes with a particular symmetry may exhibit behavior 
different from that of the general spacetime, so eventually one needs to 
tackle the general problem.  

Many of the issues involved in numerical simulations of singularities appear
in the first and simplest models studied by Berger and Moncrief: the 
Gowdy spacetimes.\cite{gowdy} These are vacuum spacetimes with two 
spacelike, commuting Killing fields that are 
orthogonally transitive.  Other systems with symmetry 
treated in this research program include the general 
vacuum $T^2$ symmetric spacetime (no orthogonal transitivity) and the 
vacuum $U(1)$ symmetric spacetimes (only a single Killing field).   
Nonetheless, in all these cases the method made use in an essential
way of the symmetry, and there was no clear way to generalize these 
methods to the case of no symmetry: a different method was needed.

This different method was provided by Uggla et al\cite{uggla}, using a set of 
scale invariant variables in a tetrad formalism.  Though this sort
of method had originally been developed to treat homogeneous 
spacetimes, it was generalized without too much difficulty to the
case with no symmetry, and it turns out to be suitable for a numerical
treatment.

In this paper, I will tell the begining and end of the story while
omitting the middle.  Section 2 will cover the Gowdy spacetimes and 
what has been learned about them from numerical simulations.  Section 3
treats the Uggla et al system and its numerical implementation.  Section
4 will consider open problems and areas for improvement in the numerical 
simulation of singularities.

\section{Gowdy Spacetimes}

The Gowdy spacetimes on ${T^3}\times R$ take the form
\begin{equation}
d{s^2} = {e^{(\lambda + \tau )/2}} ( - {e^{-2\tau}}d{\tau ^2}+ d{x^2})
+{e^{-\tau}}[{e^P}{{(dy+Qdz)}^2}+{e^{-P}}d{z^2}] 
\label{gowdymetric}
\end{equation}
where the metric functions $P, Q$ and $\lambda$ depend only on the 
time coordinate $\tau$ and the spatial coordinate $x$.  Thus 
${(\partial /\partial y)}^a$ and ${(\partial /\partial z)}^a$ 
are Killing fields.
The $T^3$ spatial topology is imposed by having 
$0 \le x,y,z \le 2\pi$ and having $P, Q$ and $\lambda$ be periodic
functions of $x$.  

The time coordinate $\tau$ has an invariant geometric
meaning, since the area of the orbits of the symmetry group is 
$4 {\pi ^2}{e^{-\tau}}$.  This area approaches zero as the singularity
is approached, and thus in terms of coordinates the singularity is at
$\tau = \infty$ (though it can certainly be reached by observers in 
finite proper time).  This property of the time coordinate, that the 
singularity is reached at infinite time, is very helpful for numerical 
simulations. A numerical simulation must stop at the earliest time 
when any part of a time slice reaches the singularity.  Thus an ideal
time coordinate is one in which no constant time slice ever hits the
singularity, but in which the slices of constant time approach arbitrarily
close to the singularity as the time goes to infinity.  In this way, 
the behavior of the spacetime as the singularity is approached can be 
read off as the limiting behavior of the spacetime at large values
of the time coordinate.   

For spacetimes of the form in eqn. (\ref{gowdymetric}) the vacuum Einstein
equations become
\begin{eqnarray}
{P_{,\tau \tau}} &=& {e^{2P}}{Q_{,\tau}^2}+{e^{-2\tau}}{P_{,xx}}-
{e^{2(P-\tau )}}{Q_{,x}^2} 
\label{gowdyP}\\
{Q_{,\tau \tau}} &=& - 2 {P_{,\tau}}{Q_{,\tau}}+{e^{-2\tau}}({Q_{,xx}}
+2{P_{,x}}{Q_{,x}}) 
\label{gowdyQ}
\end{eqnarray}
as well as equations (essentially the usual constraint equations) 
that determine $\lambda $ once $P$ and $Q$ are known.  Here the notation
for derivatives is that ${_{,a}}={\partial /\partial a}$. 

\begin{figure}
\includegraphics{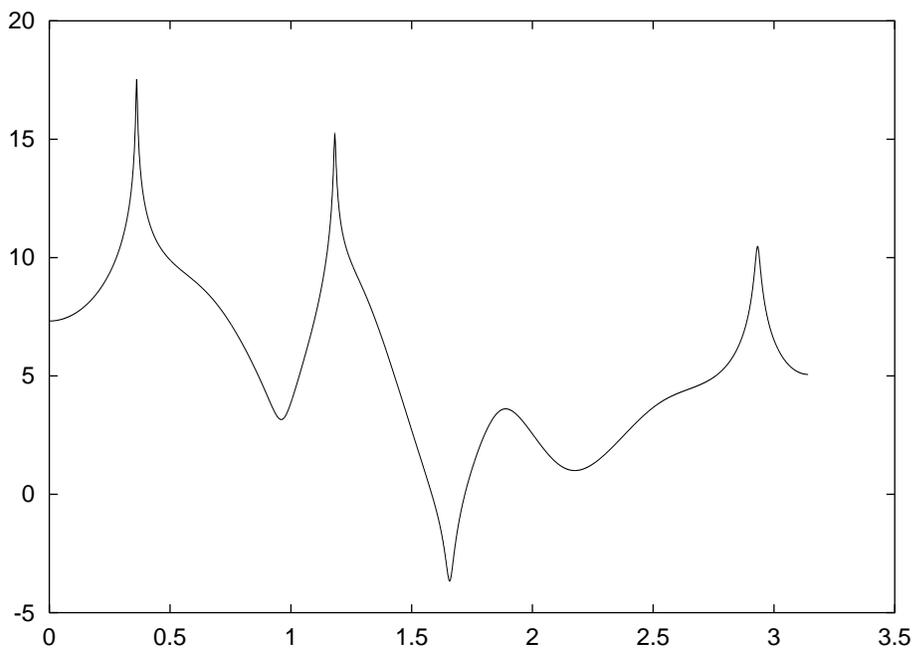}
\caption{\label{Pfig}Plot of $P$ {\it vs} $x$ for ${v_0}=5$ at
$\tau =10$. Here $0 \le x \le \pi$}
\end{figure}

\begin{figure}
\includegraphics{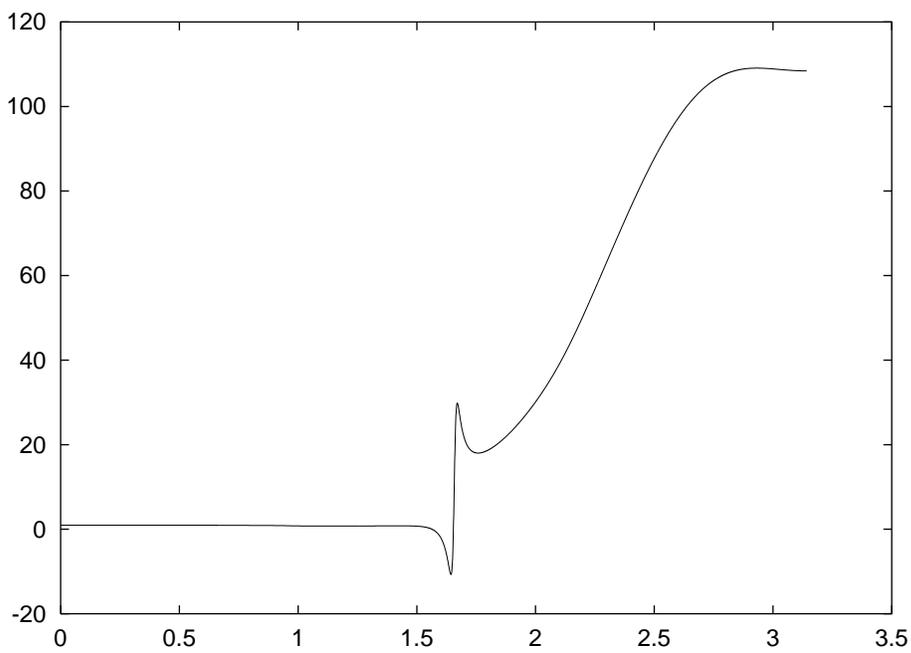}
\caption{\label{Qfig}Plot of $Q$ {\it vs} $x$ for ${v_0}=5$ at
$\tau =10$.  Here $0 \le x \le \pi$}
\end{figure}

One part of the BKL conjecture is that as the singularity is approached,
terms in the Einstein field equations containing spatial derivatives
become negligible compared to those that contain time derivatives.  A
glance at equations (\ref{gowdyP}) and (\ref{gowdyQ}) indicates how that
might come about: in these equations, all the terms that involve spatial
derivatives contain a factor of $e^{-2\tau}$.  As $\tau \to \infty$ one might
therefore expect these terms to be negligible. 
Neglecting those terms allows one to solve equations (\ref{gowdyP})
and (\ref{gowdyQ}) in closed form and leads to the expected asymptotic
behavior for large $\tau$
\begin{eqnarray}
P \to p(x)+\tau v(x)
\label{largetauP}
\\
Q \to q(x)
\label{largetauQ}
\end{eqnarray}
for some functions $p(x), \; v(x)$ and $q(x)$.  
However, a second glance
at equation (\ref{gowdyP}) reveals that things are not quite so simple:
the last term in that equation contains not the exponential of $-2\tau$ but 
instead the exponential of $2(P-\tau )$.  Thus if $P$ grows faster than
$\tau $ this term might not be negligible.  One can resolve this question
by doing a numerical simulation as was done in\cite{bv1}.  There
are many ways to perform such a simulation, and in fact the method
of \cite{bv1} uses a sophisticated technique called symplectic
integration.  However, one can just as easily use the more standard
numerical relativity technique of using centered differences for spatial
derivatives, and the Iterative Crank-Nicholson (ICN) method for time 
derivatives (after first rewriting the system as one that is first order
in time).  The results of such a simulation are given in figures 
\ref{Pfig} and \ref{Qfig}.  

Here the initial data at $\tau =0$ are 
$P=0, \; {P_{,\tau}}=5 \cos x, \; Q = \cos x , \; {Q_{,\tau}}=0$
and the system has been evolved to $\tau = 10$.     
Note the presence of narrow features in these graphs that have been 
termed ``spikes.'' Away from the spikes, the asymptotic form of the
solution is that of equations (\ref{largetauP}) and (\ref{largetauQ}) 
with $v(x)<1$.
However, at 
the upward pointing spikes in $P$, we have that $P$ grows faster
than $\tau$ and that the spike becomes ever narrower as the singularity is
approached.  Thus the simulations do show that as the singularity is approached
the dynamics becomes one where the spatial derivatives can be neglected; 
however, they also show that there are special points where the dynamics
is different from the neighboring points.  This is a challenge for 
the numerics even for a 1+1 dimensional
simulation.  One can use adaptive mesh refinement to resolve the 
spikes\cite{stewart} or simply use a very large number of spatial points;
but the spikes narrow exponentially in time and thus will eventually become
too narrow for the numerical method to resolve them.     

\section{scale invariant tetrad systems}

The Gowdy spacetimes have two features that are very convenient for the
study of singularities: the first is the time coordinate that goes
to infinity as the singularity is approached.  The second is that 
even though the metric components themselves are exponential functions
of the time coordinate, the quantities $P$ and $Q$ that are evolved by
the numerics have much more mild behavior, growing no more than linearly
with time.  One would like to have both of these features for a system
that has no symmetries whatsoever.  It turns out that a convenient
approach came from the study of homogeneous spacetimes using the 
tetrad formalism.  In a homogeneous spacetime, each time slice has
constant mean curvature, and while various geometric quantities blow
up as the singularity is approached, they do so as particular powers
of the mean curvature.  Thus, one can introduce scale invariant
variables by dividing each geometric quantity in the equations of motion
by the appropriate power of mean curvature and then rewriting the field
equations in terms of these scale invariant quantities.

Not only does this work for homogeneous spacetimes; but it can also be
generalized to the case of no symmetry at all, as was done by Uggla
et al\cite{uggla}.  The system of reference \cite{uggla} is the following:
the spacetime
is described in terms of a coordinate system ($t,{x^i}$) and a tetrad
(${{\bf e}_0},{{\bf e}_\alpha}$)  where both the spatial coordinate
index $i$ and
the spatial tetrad index $\alpha $ go from 1 to 3.  Choose
${\bf e}_0$ to be hypersurface orthogonal with the relation between
tetrad and coordinates of the form
${{\bf e}_0} = {N^{-1}}{\partial _t}$ and
${{\bf e}_\alpha} =
{{e_\alpha }^i}{\partial _i}$
where $N$ is the lapse and the shift is chosen to be zero.
Choose the spatial frame $\{ {{\bf e}_\alpha} \}$ to be
Fermi propagated along the integral curves of ${\bf e}_0$.
The commutators of the tetrad components are decomposed as follows:
\begin{eqnarray}
[{{\bf e}_0},{{\bf e}_\alpha}] &=& {{\dot u}_\alpha}{{\bf e}_0}
-(H {{\delta _\alpha}^\beta}
+{{\sigma _\alpha}^\beta})
{{\bf e}_\beta}
\\
\left [ {{\bf e}_\alpha },{{\bf e}_\beta} \right ]  &=&
(2 {a_{[\alpha}}{{\delta _{\beta ]}}^\gamma}
+ {\epsilon _{\alpha \beta \delta }}{n^{\delta \gamma}}){{\bf e}_\gamma}
\end{eqnarray}
where $n^{\alpha \beta}$ is symmetric, and $\sigma ^{\alpha \beta}$ is
symmetric and trace free.
Scale invariant variables are defined as follows:
$\{ {\parb_0},{\parb_\alpha} \} \equiv \{
{{\bf e}_0},{{\bf e}_\alpha} \} /H$
\begin{equation}
\{ {{E_\alpha}^i}, {\Sigma _{\alpha \beta }}, {A^\alpha} ,
{N_{\alpha \beta }} \} \equiv \{ {{e_\alpha}^i} ,
{\sigma _{\alpha \beta }} , {a^\alpha}, {n_{\alpha \beta}} \} /H
\end{equation}
$q+1 \equiv - {\parb _0} \ln H $ and
${r_\alpha} \equiv - {\parb _\alpha} \ln H$.

Finally choose the lapse to be $N={H^{-1}}$.  In the conventions of 
\cite{uggla} the singularity is in the negative $t$ direction and one
expects it to be approached as $t \to -\infty$. The relation between
scale invariant frame derivatives and coordinate derivatives is
${\parb _0} ={\partial _t}$ and
${\parb _\alpha} = {{E_\alpha }^i} {\partial _i}$.
From the vacuum Einstein equations one obtains the following
evolution equations:
\begin{eqnarray}
{\partial _t} {{E_\alpha}^i} &=& {{F_\alpha}^\beta}{{E_\beta}^i}
\label{ev1}
\\
{\partial _t} {r_\alpha} &=& {{F_\alpha}^\beta}{r_\beta}+{\parb _\alpha}q
\label{ev2}\\
{\partial _t} {A^\alpha} &=& {{F^\alpha}_\beta}{A^\beta}+
{\textstyle \frac 1 2}{\parb  _\beta}{\Sigma ^{\alpha \beta}}
\label{ev3}
\\
\nonumber
{\partial _t} {\Sigma ^{\alpha \beta}} &=& (q-2) {\Sigma ^{\alpha \beta}}
- 2 {{N^{<\alpha}}_\gamma}{N^{\beta > \gamma}} + {{N_\gamma }^\gamma}
{N^{<\alpha \beta >}}
+ {\parb ^{<\alpha}}{r^{\beta >}}
\\
&-& {\parb ^{<\alpha}}{A^{\beta >}}
+ 2{r^{<\alpha }}{A^{\beta >}}
+ {\epsilon ^{\gamma \delta < \alpha}}({\parb _\gamma } - 2 {A_\gamma})
{{N^{\beta > }}_\delta}
\label{ev4}
\\
{\partial _t}{N^{\alpha \beta}} &=& q {N^{\alpha \beta }} +
2 {{\Sigma ^{(\alpha }}_\delta}{N^{\beta ) \delta }} -
{\epsilon ^{\gamma \delta (\alpha }}{\parb _\gamma }
{{\Sigma ^{\beta )}}_\delta}
\label{ev5}
\\
\nonumber
{\partial _t} q &=& \left [ 2 (q-2) + {\textstyle \frac 1 3}
\left ( 2 {A^\alpha } - {r^\alpha}\right ) {\parb _\alpha}
- {\textstyle \frac 1 3} {\parb ^\alpha}{\parb _\alpha}
\right ] q
\\
&-& {\textstyle \frac 4 3} {\parb _\alpha}{r^\alpha} +
{\textstyle \frac 8 3}{A^\alpha}{r_\alpha} + {\textstyle \frac 2 3}
{r_\beta}{\parb _\alpha}{\Sigma ^{\alpha \beta}}
- 2 {\Sigma ^{\alpha \beta}}{W_{\alpha \beta}}
\label{ev6}
\end{eqnarray}
Here angle brackets denote the symmetric trace-free part, and
$F_{\alpha \beta }$ and $W_{\alpha \beta} $ are given by
\begin{eqnarray}
{F_{\alpha \beta }} &\equiv & q {\delta _{\alpha \beta}} - {\Sigma _{\alpha \beta}}
\\
\nonumber
{W_{\alpha \beta }} &\equiv & {\textstyle \frac 2 3}{N_{\alpha \gamma}}
{{N_\beta}^\gamma}
- {\textstyle \frac 1 3} {{N^\gamma }_\gamma}
{N_{\alpha \beta }}
+ {\textstyle \frac 1 3} {\parb _\alpha}
{A_\beta}
\\
&-& {\textstyle \frac 2 3} {\parb _\alpha} {r_\beta}
- {\textstyle \frac 1 3}
{{\epsilon ^{\gamma \delta }} _\alpha } \left ( {\parb _\gamma}
- 2 {A_\gamma}\right ) {N_{\beta \delta}}
\end{eqnarray}
In addition to the evolution equations, the variables are subject
to the vanishing of the following constraints: 
\begin{eqnarray}
{{({{\cal C}_{\rm com}})}^i _{\alpha \beta}} &\equiv &
2 ( {\parb _{[\alpha }} - {r_{[\alpha}}-{A_{[\alpha}}){{E_{\beta ]}}^i}
- {\epsilon _{\alpha \beta \delta}}{N^{\delta \gamma}}{{E_\gamma }^i}
\label{cn1}
\\
\nonumber
{{\cal C}_{\rm G}} &\equiv & 1 + {\textstyle \frac 1 3}
(2 {\parb _\alpha} - 2 {r_\alpha} - 3 {A_\alpha}){A^\alpha} -
{\textstyle \frac 1 6}{N_{\alpha \beta}}{N^{\alpha \beta}}
\\
&+&{\textstyle \frac 1 {12}} {{({{N^\alpha}_\alpha})}^2} -
{\textstyle \frac 1 6} {\Sigma _{\alpha \beta}}{\Sigma ^{\alpha \beta}}
\label{cn2}
\\
{{({{\cal C}_{\rm C}})}^\alpha} &\equiv &{\parb _\beta}
{\Sigma ^{\alpha \beta}}+ 2 {r^\alpha} - {{\Sigma ^\alpha}_\beta}
{r^\beta} - 3 {A_\beta}{\Sigma ^{\alpha \beta}}
-{\epsilon ^{\alpha \beta \gamma}}{N_{\beta \delta}}
{{\Sigma _\gamma}^\delta}
\label{cn3}
\\
{{\cal C}_q} &\equiv & q - {\textstyle \frac 1 3} {\Sigma ^{\alpha \beta}}
{\Sigma _{\alpha \beta}}+{\textstyle \frac 1 3} {\parb _\alpha }
{r^\alpha} - {\textstyle \frac 2 3}{A_\alpha}{r^\alpha}
\label{cn4}
\\
{{({{\cal C}_{\rm J}})}^\alpha} &\equiv & ({\parb _\beta} - {r_\beta})
({N^{\alpha \beta}} + {\epsilon ^{\alpha \beta \gamma }}{A_\gamma})
- 2 {A_\beta}{N^{\alpha \beta}}
\label{cn5}
\\
{{({{\cal C}_{\rm W}})}^\alpha} &\equiv & [ {\epsilon ^{\alpha \beta
\gamma }}({\parb _\beta} - {A_\beta}) - {N^{\alpha \gamma}}]{r_\gamma}
\label{cn6}
\end{eqnarray} 
In fact, this set of equations yields a class of evolution equations, 
since one can always add any of the constraint equations to the right hand
side of any of the evolution equations.  In fact, the system as written
is not well posed, but it becomes well posed if one adds $d$ times
equation (\ref{cn3}) to the right hand side of equation (\ref{ev3}) 
where $d$ is any number 
less then $-1/2$.\cite{meandcarsten}

The properties of the time coordinate come about in the choice of lapse:
$N={H^{-1}}$ where $3H$ is the mean curvature.  This is sometimes known
as inverse mean curvature flow: a system that has been studied in the
context of Riemannian geometry.  For our purposes, the main point is that
it is expected that the mean curvature blows up as the singularity
is approached.  By using inverse mean curvature flow, the lapse gets
smaller the nearer we approach the singularity.  Inverse mean curvature
flow is a parabolic equation, as can be seen in our case by the
presence of second spatial derivatives on the right hand side of equation
(\ref{ev6}).  In fact, this is the only place where second derivatives 
occur, and the entire evolution is of a mixed hyperbolic-parabolic 
nature.\cite{meandcarsten} The other nice property of the system has to
do with the use of scale invariant variables: thus for example, 
as the singularity is
approached, one would expect both the mean curvature $3H$ and 
the shear $\sigma_{\alpha \beta}$ 
to blow up, but the scale invariant quantity
${\Sigma _{\alpha \beta}}={\sigma _{\alpha \beta}}/H$ to remain finite.     
 
Nonetheless, a mixed hyperbolic-parabolic system is somewhat unusual 
from the point of view of numerical relativity.  Essentially it comes
about because while we expect gravity to be described by a hyperbolic
system, we always have a choice of gauge, and in this particular case
we have chosen one that is parabolic.  This is somewhat analogous to 
a more usual situation in numerical relativity where one chooses a gauge
like maximal slicing which is an elliptic condition, and the
vacuum Einstein equation then becomes a mixed hyperbolic-elliptic system.  
Lars Andersson has suggested that one could study singularities by 
simply using the system of ref. \cite{uggla} but with constant mean curvature
time slices as the gauge instead of inverse mean curvature flow.\cite{lars}
To implement this suggestion, we simply choose the time coordinate
$t$ so that $3H={e^{-t}}$. (Note that the singularity is approached
as $t \to -\infty$). This gives rise to the following elliptic
equation
\begin{equation}
-{\parb ^\alpha}{\parb _\alpha}{\cal N} + 2 {A^\alpha}{\parb _\alpha}{\cal N}
+(3+{\Sigma ^{\alpha \beta}}{\Sigma _{\alpha \beta}}){\cal N}=3
\label{lapse}
\end{equation} 
where the scale invariant lapse $\cal N$ is defined by ${\cal N}=NH$.
Since $H$ is now constant on the surfaces of constant time, it follows
that ${r_\alpha} = 0$.  Therefore, equation (\ref{cn4}) simply becomes
$q=(1/3){\Sigma ^{\alpha \beta}}{\Sigma _{\alpha \beta}}$.  Thus, the only
quantities that we need to evolve are 
${{E_\alpha }^i}, \; {A_\alpha}, \; {N^{\alpha \beta}}$ and 
$\Sigma _{\alpha \beta}$.  These quantities are subject to the following
evolution equations.
\begin{eqnarray}
{\partial _t} {{E_\alpha}^i} &=& {{E_\alpha}^i} - {\cal N} (
{{E_\alpha}^i} + {{\Sigma _\alpha}^\beta}{{E_\beta}^i})
\\
{\partial _t} {A_\alpha} &=& {A_\alpha} + ({\textstyle {1\over 2}}
{{\Sigma _\alpha}^\beta}-{{\delta _\alpha}^\beta}){\parb _\beta}{\cal N}
+{\cal N}\left ( {\textstyle {1\over 2}} {\parb _\beta} 
{{\Sigma _\alpha}^\beta} - {A_\alpha}
-{{\Sigma _\alpha}^\beta}{A_\beta}\right )
\\
\nonumber
{\partial _t}{N^{\alpha \beta}} &=& {N^{\alpha \beta}} - 
{\epsilon ^{\gamma \delta (\alpha}}{{\Sigma _\delta}^{\beta )}}
{\parb _\gamma}{\cal N} 
\\
&+& {\cal N}\left ( - {N^{\alpha \beta}} + 2
{{N^{(\alpha}}_\lambda}{\Sigma ^{\beta )\lambda}} -
{\epsilon ^{\gamma \delta (\alpha}}{\parb _\gamma}
{{\Sigma _\delta}^{\beta )}}\right )  
\\
\nonumber
{\partial _t} {\Sigma _{\alpha \beta}} &=&  {\Sigma _{\alpha \beta}} 
+ {\parb _{<\alpha}}{\parb _{\beta >}}{\cal N} +{A_{<\alpha}}
{\parb _{\beta >}} {\cal N} + {\epsilon _{\gamma \delta (\alpha}}
{{N_{\beta )}}^\delta}{\parb ^\gamma }{\cal N}
\\
\nonumber
&+& {\cal N} \bigl [ - 3 {\Sigma _{\alpha \beta}} - {\parb _{<\alpha}}
{A_{\beta >}} - 2 {{N_{<\alpha}}^\gamma}{N_{\beta >\gamma}}
+ {{N^\gamma}_\gamma}{N_{<\alpha \beta >}}
\\
&+&{\epsilon _{\gamma \delta (\alpha}} ( 
{\parb ^\gamma}{{N_{\beta )}}^\delta}   
- 2 {A^\gamma}{{N_{\beta )}}^\delta} ) \bigr ]
\end{eqnarray}
and to the vanishing of the following constraints.
\begin{eqnarray}
{{({{\cal C}_{\rm com}})}^{\lambda i}} &\equiv & 
{\epsilon ^{\alpha \beta \lambda}} [ {\parb _\alpha}{{E_\beta}^i} - 
 {A_\alpha}{{E_\beta}^i} ] - {N^{\lambda \gamma}}{{E_\gamma }^i}
\\
{{({{\cal C}_{\rm J}})}^\gamma} &\equiv & {\parb _\alpha} 
{N^{\alpha \gamma}} + {\epsilon ^{\alpha \beta \gamma }}
{\parb _\alpha }{A_\beta}
- 2 {A_\alpha}{N^{\alpha \gamma}}
\\
{{({{\cal C}_{\rm C}})}_\alpha} &\equiv & {\parb _\beta}
{{\Sigma _\alpha} ^\beta} - 3 {{\Sigma _\alpha} ^\beta} {A_\beta} 
-{\epsilon _{\alpha \beta \gamma}}{N^{\beta \delta}}
{{\Sigma _\delta}^\gamma}
\\
\nonumber
{{\cal C}_{\rm G}} &\equiv & 1 + {\textstyle \frac 2 3}
{\parb _\alpha}{A^\alpha}  -  {A_\alpha}{A^\alpha} -
{\textstyle \frac 1 6}{N_{\alpha \beta}}{N^{\alpha \beta}}
\\
&+&{\textstyle \frac 1 {12}} {{({{N^\alpha}_\alpha})}^2} -
{\textstyle \frac 1 6} {\Sigma _{\alpha \beta}}{\Sigma ^{\alpha \beta}}
\end{eqnarray}
To distinguish these two versions of the scale invariant tetrad system,
we will call the first one the inverse mean curvature system and the
second one the constant mean curvature (or just CMC) system.

In order to evolve either one of these scale invariant systems, one
must find initial data that satisfies the constraint equations.  This
is done essentially by using the York method to find initial data for
the usual spatial metric and extrinsic curvature variables and then
translating that initial data into data for the scale invariant tetrad
variables.  In particular, choose spatial topology $T^3$, constant
mean curvature, and choose
the spatial metric to be of the form ${h_{ab}} = {\psi ^4} {\delta _{ab}}$
where $\delta _{ab}$ is a flat metric.  Then, the momentum constraint 
becomes a differential equation in the flat space coordinates for the 
trace-free part of the extrinsic curvature (multiplied by a power of
the conformal factor), which on $T^3$ with a conformally
flat metric simply becomes an algebraic equation for the Fourier components
of that quantity.  The Hamiltonian 
constraint then becomes an elliptic equation for the conformal factor
$\psi$.  A simple family of initial data of this kind is the following:
$H={\rm constant}, \; {N^{\alpha \beta}}=0, \; 
{{E_\alpha}^i}=({\psi ^{-2}}/H) {{\delta _\alpha}^i},
{A_\alpha} = - 2 {\psi ^{-1}}{\parb _\alpha} \psi $
and ${\Sigma _{\alpha \beta}} = ( - {\psi ^{-6}}/H) {\rm diag} (
{{\bar \Sigma}_1}, {{\bar \Sigma}_2}, {{\bar \Sigma}_3})$.
with $\psi$ satisfying the equation 
\begin{equation}
{\partial ^i} {\partial _i} \psi = {\textstyle \frac 1 8}  (
6 {H^2} {\psi ^5} - {{\bar \Sigma}^i}{{\bar \Sigma}_i}{\psi ^{-7}} )
\label{elliptic}
\end{equation}
and the ${\bar \Sigma}_i$ given by
\begin{eqnarray}
\nonumber
{{\bar \Sigma}_1}&=&{a_2}\cos y + {a_3} \cos z +{b_2}+{b_3}
\\
\nonumber
{{\bar \Sigma}_2}&=&{a_1} \cos x - {a_3} \cos z +{b_1}-{b_3}
\\
{{\bar \Sigma}_3} &=& -{a_1} \cos x - {a_2} \cos y - {b_1} - {b_2}
\label{sigmabar}
\end{eqnarray}
with the $a_i$ and $b_i$ constants.  This set of conditions is enough
to satisfy the constraints of the CMC system.  
If one wants to satisfy the constraints
of the inverse mean curvature system, one must add the condition
${r_\alpha}=0$ and one must solve equation (\ref{cn4}) for $q$.  

We now consider the form of the BKL conjecture for these scale invariant
systems.  First note that all spatial derivatives occur in the form
${\parb _\alpha}={{E_\alpha}^i}{\partial _i}$.  Thus one would expect
spatial derivatives to be negligible as the singularity is approached
provided that ${{E_\alpha}^i}$ becomes negligible and provided that 
any derivatives with respect to spatial coordinates at least do not
become so large to overwhelm the smallness of ${{E_\alpha}^i}$.  From 
the form of the evolution equations, there are reasons to expect this 
to occur,\cite{dgprl} and these reasons also lead one to expect 
that $A_\alpha$
(and in the inverse mean curvature case also $r_\alpha$) become negligibly
small.  However, the ultimate test for all this is numerical: evolve 
initial data towards the singularity and see whether these quantities 
do indeed become negligible.  

Note that as in the Gowdy case spatial derivatives becoming negligible 
does {\it not} mean that the spacetime is becoming homogeneous.  Instead,
the dependence of the metric components on space can become quite large
(especially at spikes); however since in the field equations these spatial 
derivatives are
multiplied by a small quantity ($e^{-2\tau}$ in the Gowdy case and
${E_\alpha}^i$ in the scale invariant tetrad case) this means that the
spatial derivatives make a negligible contribution to the field equations.
Nonetheless, with spatial derivatives becoming negligible in the field
equations, this means that the dynamics at each spatial point is that
of a homogeneous spacetime (albeit a different homogeneous spacetime for 
each spatial point).  Since there are many different homogeneous spacetimes
(classified according to the Bianchi classification) there remains the
question of {\it which} homogeneous spacetimes describe the dynamics of
the general singularity.  Another part of the BKL conjecture is that
it is the dynamics of the Bianchi type IX spacetimes, also called
Mixmaster dynamics, that describes the
approach to the singularity.  This dynamics consists of ``epochs'' where
the components of ${\Sigma}_{\alpha \beta}$ are approximately constant,
punctuated by short ``bounces'' where they change rapidly.  The components
of $N_{\alpha \beta}$ are negligible during the epochs and exhibit rapid
growth and equally rapid decay during the bounces.  All this should also
be checked numerically. However if the spatial
derivative terms, as well as $A_\alpha$ (and $r_\alpha$ in the inverse 
mean curvature case) are negligible, then one can replace the evolution
equations and constraint equations with a set of truncated equations where 
these terms are absent.  An analysis of these truncated equations\cite{dgprl}
(which due to the neglect of the spatial derivatives are ODEs) shows that 
this sort of local Mixmaster dynamics is exactly what is to be
expected.   Thus the main thing to be checked from the numerics is whether
the terms that are expected to become negligible do in fact become negligible
as the singularity is approached. 

We now turn to numerical methods and presentation of numerical
results.  The results for the inverse mean curvature system have
been presented in\cite{dgprl} so here we will concentrate on the CMC
system. The spatial topology is chosen to be $T^3$ with coordinates
$(x,y,z)$ satisfying $0 \le x,y,z \le 2 \pi$ and with $0$ and $2\pi$
identified.  This is implemented numerically by using $n+2$ grid points
in each spatial direction with spacing $dx=2\pi /n$ between adjacent grid
points.  The variables on gridpoints $2$ to $n+1$ are evolved, while those
at gridpoints $1$ and $n+2$ are set using the periodic boundary conditions.
Derivatives with respect to spatial coordinates are evaluated using 
centered differences, while time derivatives are implemented using the
ICN method.  Equation (\ref{lapse}) for the scale invariant lapse is 
solved at each time step using a multigrid method.  The initial data is
of the form in equation (\ref{sigmabar}) with equation (\ref{elliptic})
for the conformal factor solved by relaxation.   
   
\begin{figure}
\includegraphics{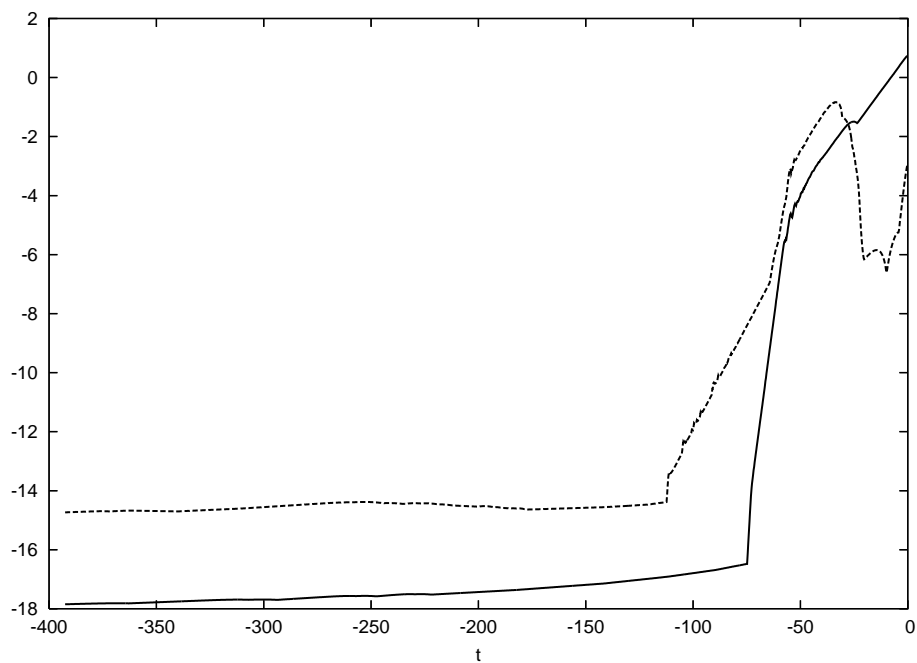}
\caption{\label{lnmaxfig}Plot of the maximum values of 
$\ln |{{E_\alpha}^i}|$ (solid line) and $\ln |{A_\alpha}|$ (dotted line) 
{\it vs} $t$} 
\end{figure}

\begin{figure}
\includegraphics{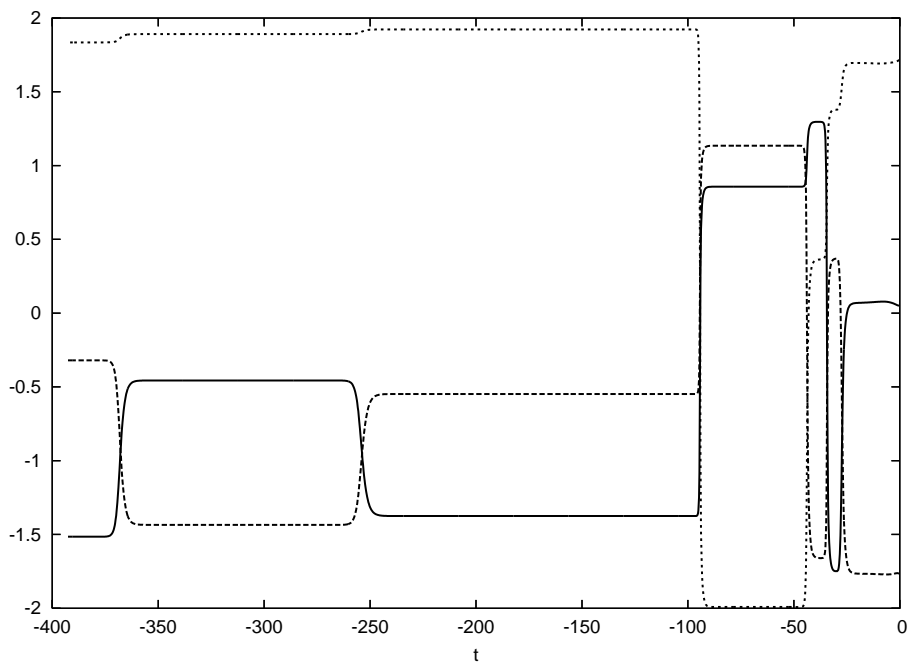}
\caption{\label{sigmafig}Plot of the components of $\Sigma _{\alpha \beta}$ 
{\it vs} $t$}
\end{figure}

\begin{figure}
\includegraphics{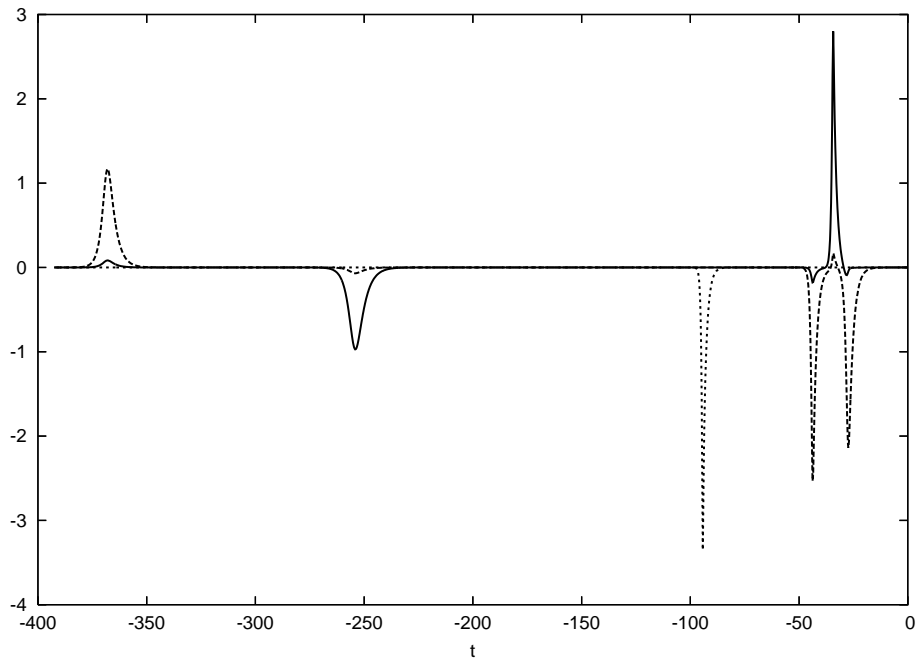}
\caption{\label{nfig}Plot of the components of $N_{\alpha \beta}$ 
{\it vs} $t$}
\end{figure}

The simulation was done on a SunBlade 2000 with $n=32$ and 
with $H=1/3,\; {a_i}=(0.2,0.1,0.02)$ and 
${b_i}=(1.8,-0.15,0)$.  Figure \ref{lnmaxfig} shows for each time
the logarithms of the maximum values 
(over all spatial points and all indicies) of
$|{{E_\alpha}^i}|$ and $|{A_\alpha}|$.  Note that these quantities become 
negligibly small and so the BKL conjecture is supported.  Since spatial
derivatives are negligibly small, the dynamics is local and can be
studied by simply looking at the behavior with time of the scale invariant
quantities at a single spatial point.  Figures \ref{sigmafig} and 
\ref{nfig} show this sort
of behavior for the diagonal components of $\Sigma_{\alpha \beta}$ and
$N_{\alpha \beta}$.  
\begin{figure}
\includegraphics{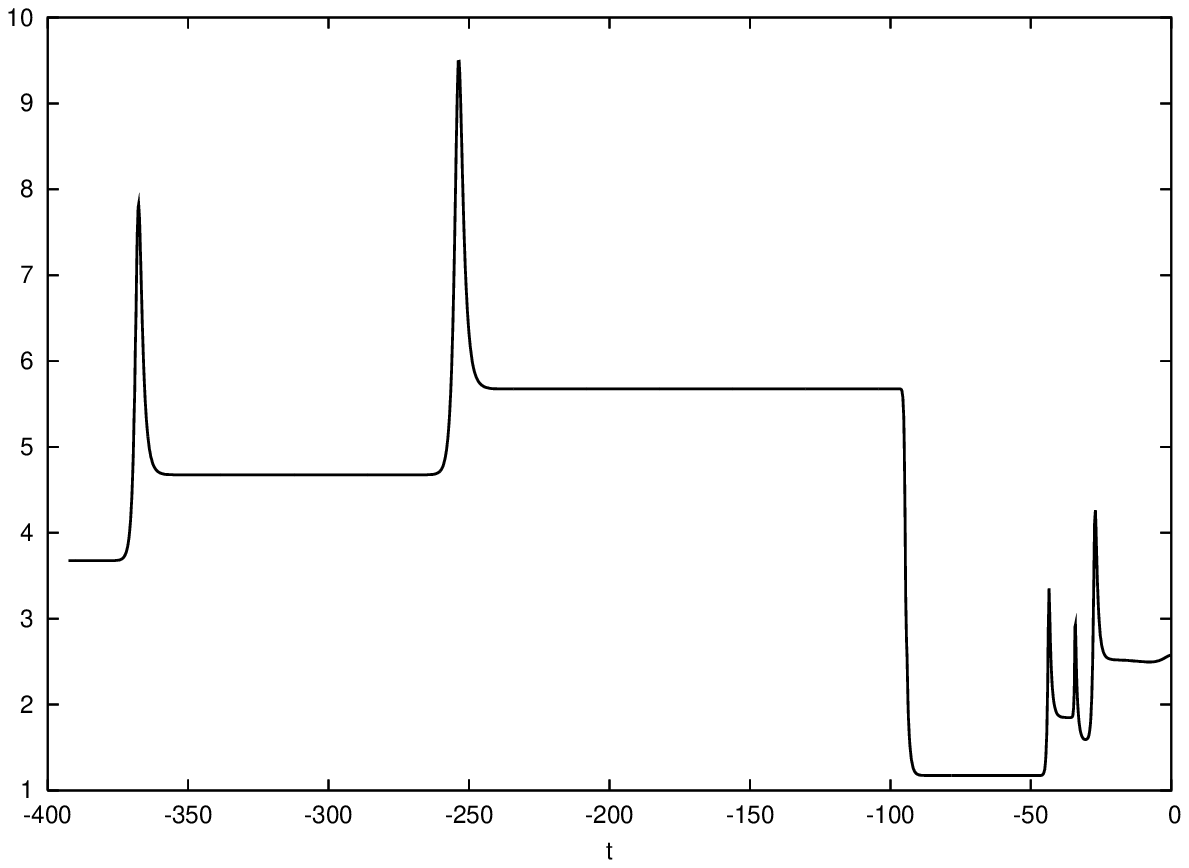}
\caption{\label{ufig}Plot of $u$ {\it vs} $t$} 
\end{figure}
Note that sufficiently near the singularity, 
the behavior is a series of epochs where the components of 
$\Sigma _{\alpha \beta}$ are approximately constant, punctuated by bounces
where the components of $\Sigma _{\alpha \beta}$ change rapidly.  Also
note that $N_{\alpha \beta}$ is negligible except at the bounces.  The
tensor $\Sigma _{\alpha \beta}$ has three eigenvalues.  However, 
$\Sigma _{\alpha \beta}$ is traceless, and during each epoch satisfies
${\Sigma ^{\alpha \beta}}{\Sigma _{\alpha \beta}}=6$.  The eigenvalues of 
$\Sigma _{\alpha \beta}$ can therefore be determined by a single 
quantity $u$ defined by
\begin{equation}
{{\Sigma ^\alpha}_\beta}{{\Sigma ^\beta}_\gamma}{{\Sigma ^\gamma}_\alpha}
= 6 - {{81{u^2}{{(1+u)}^2}}\over{{(1+u+{u^2})}^3}} 
\end{equation}
(There is a unique $u \ge 1$ provided that the quantity on the left hand side 
of the equation is between $-6$ and $6$).  It is a consequence of Mixmaster 
dynamics that $u$ is constant in each epoch and that  
from one epoch to the next $u$ changes by
the map
$ u \to u - 1$ if $u \ge 2$ and $u \to 1/(u-1)$ if $1 < u \le 2$.  figure
\ref{ufig} shows the value of $u$ at the same spatial point.  
For times sufficiently near the singularity the sequence of 
$u$ values satisfies this map.

\section{open problems}

The main limitation of the results just presented is the  low resolution
of the simulation (though simulations of this sort of system but assuming
symmetry have been done with high resolution\cite{lars2}).  
This is especially true because the general singularity
has spikes analogous to those of the Gowdy spacetime.  In the system of 
scale invariant tetrad variables, these occur because of the nature of the 
bounces: a bounce occurs because in any epoch one component of 
$N_{\alpha \beta}$ grows exponentially with time until it becomes large enough
to cause a sudden change in the components of $\Sigma _{\alpha \beta}$.  This 
occurs whether that component of $N_{\alpha \beta}$ is positive or negative; 
but if it is zero then it remains zero and there is no bounce.   
Thus in general spikes occur on surfaces of co-dimension 1, so what were 
isolated points in the Gowdy case are isolated surfaces in the general case.
To properly treat the spikes in the general case will require a singularity
simulating code that is parallel and uses adaptive mesh refinement.  
Development of such a code is presently work in progress.

Another limitation of the current results is that they are only for the 
vacuum case.  Another part of the BKL conjecture is that as the singularity 
is approached, matter makes a negligible contribution to the gravitational
field equations unless it is a stiff fluid (equation of state $P=\rho$).  
In the stiff fluid case, BKL conjecture that there is a last bounce after 
which the components of $\Sigma _{\alpha \beta}$ remain essentially 
constant all the way to the singularity.  This stiff fluid part of the 
BKL conjecture is supported by a theorem of Rendall and 
Andersson\cite{larsandalan} as well as by numerical simulations.\cite{josh}  
What is
needed is some numerical simulations to cover the non-stiff fluid case.  
In \cite{uggla} scale invariant tetrad equations are given 
for fluids with equation
of state $P=k\rho$ for constant $k$.  What is needed is a numerical 
simulation of these equations to see whether for $k<1$ the fluid has
a negilgible influence on the gravitational degrees of freedom as the
singualarity is approached.  Since fluids with $k<1$ can form shocks, 
such a numerical simulation would need a high resolution shock capturing
method.

Another limitation of the current results is that they are for spacetimes
with compact spatial topology and therefore do not directly address
the physically relevant case of gravitational collapse in an asymptotically
flat spacetime. This is not as bad as it seems: since the simulations
show that the formation of singularities in spacetimes 
with $T^3$ spatial topology is essentially a local process, this means
that one can expect at least parts of the singularities produced in 
asymptotically flat spacetimes to have the same features as the 
singuarities in these simulations.  However, that still leaves open 
the possibility that for gravitational collapse in asymptotically
flat spacetimes other parts of the singularity have very different
properties.  In particularly Poisson and Israel have conjectured that
the part of the singularity near the horizon is not of the BKL type 
but is instead null rather than spacelike and with 
weaker tidal forces.\cite{eric}
Numerical simulations in the spherically symmetric case support this 
conjecture.\cite{brady,piran}  What is needed now is 
numerical simulations that don't
assume spherical symmetry.

\ack

I would like to thank Lars Andersson for helpful discussions, and
especially for the suggestion of modifying the Uggla et al system
by using CMC slices.
This work was supported by NSF grant PHY-0456655 
to Oakland University.

\end{document}